# Electrodynamics in Superconductors Explained by Proca Equations

M. Tajmar

Space Propulsion and Advanced Concepts, Austrian Research Centers GmbH – ARC

A-2444 Seibersdorf, Austria

Abstract

A fully consistent model to study electrodynamics for superconductors in the stationary and non-stationary regime has been developed based on Proca equations and a massive photon. In particular, this approach has been applied to study the electric field penetration depth in superconductors. The model shows a deviation from the charge contribution to an internal electric field compared to previous approaches.

## 1. INTRODUCTION

There is a recent discussion in the literature about how electric fields penetrate into superconductors. The London brothers originally proposed a set of equations that allowed the penetration of an electric field within a penetration depth from the surface[1]. However, as H. London failed to detect such a field experimentally[2], F. London modified the theory and discarded the equation this equation in his final work[3]. It was then assumed that electric fields penetrate superconductors similar to normal metals. In this case, electric fields are effectively shielded over the Thomas-Fermi length which is typically on the order of one Å. Based on the concept of hole superconductivity, Hirsch recently proposed that a superconductor should have an outside pointing electric field in its interior[4], using again the London brothers idea that non-zero electric fields can exist within the penetration depth that is much larger than the Thomas-Fermi screening length[5]. Although this approach was challenged by Koyama[6], a number of experiments indeed show that the electric field penetration depth is different in a superconductor compared to a normal conductor. Jenks et al[7] performed such measurements on an YBCO High-$T_c$ superconductor and reported that the penetration depth down to $T_c$ was decreasing with temperature as expected from the Thomas-Fermi theory, but once the YBCO was superconductive, the electric field penetration depth increased again with temperature. Tao et al[8-10] recently observed the formation of balls when low and High-Tc superconducting particles were exposed to a strong electric field. In their analysis, they had to assume that the electric field penetration depth is at least an order of magnitude higher than the Thomas-Fermi length[9].

For the case of rotating superconductors, electric fields are predicted inside the superconductor to counterbalance the centrifugal force. Discrepancies regarding the sign of this electric field have been found in some publications[11-13].

This paper derives the electrodynamics of a superconductor in stationary and non-stationary conditions using the well-established approach from quantum field theory by applying Proca equations[14]. The London equations as well as other effects such as the London moment or electric fields from rotating superconductors will be derived. The results from this model seem to confirm some of the assumptions regarding electric field screening of superconductors and they provide an alternative interpretation of well-known effects such as the Meissner effect and the London moment that can answer some of the criticism raised in standard London theory[13].

## 2. ELECTRODYNAMICS IN STATIONARY SUPERCONDUCTORS

In quantum field theory, superconductivity is explained by a massive photon, which acquired mass due to gauge symmetry breaking and the Higgs mechanism[14]. The wavelength of the photon is interpreted as the London penetration depth. With a non-zero photon mass, the usual Maxwell equations transform into the so-called Proca equations which will form the basis for our assessment in superconductors and are only valid for the superconducting electrons[15]:

$$\nabla \cdot \vec{E} = \frac{\rho}{\varepsilon_0} - \frac{1}{\lambda_L^2} \phi \qquad (1)$$

$$\nabla \cdot \vec{B} = 0 \qquad (2)$$

$$\nabla \times \vec{E} = -\frac{\partial \vec{B}}{\partial t} \qquad (3)$$

$$\nabla \times \vec{B} = \mu_0 \vec{j}_s + \frac{1}{c^2}\frac{\partial \vec{E}}{\partial t} - \frac{1}{\lambda_L^2}\vec{A} \quad . \qquad (4)$$

The Meissner effect is obtained by taking the curl of Equ. (4). Omitting the radiation term leads to

$$\Delta \vec{B} = -\mu_0 (\nabla \times \vec{j}_s) + \frac{1}{\lambda_L^2}\vec{B} \quad . \qquad (5)$$

Contrary to the usual London interpretation, in the Proca model we don't need a shielding current flowing as the magnetic field penetration is screened by the massive photon. This can be also understood from the rigidity of the Cooper-pair superfluid which was frozen into its initial state during cooling down. With no current flowing, Equ. (5) gives the usual Meissner effect

$$\Delta \vec{B} = \frac{1}{\lambda_L^2}\vec{B} \quad . \qquad (6)$$

Without the knowledge of a massive photon, we have to assume a shielding current of

$$\nabla \times \vec{j}_s = -\frac{1}{\mu_0 \lambda_L^2}\vec{B} \qquad (7)$$

in standard Maxwell equations to arrive at the same result, which is the 2. London equation. The time derivative together with the induction law of Equ. (3) yields the 1. London equation

$$\frac{\partial \vec{j}_s}{\partial t} = \frac{1}{\mu_0 \lambda_L^2}\vec{E} \quad . \qquad (8)$$

Now we will do the same analysis for the electric field. By taking the gradient of Equ. (1), we arrive at

$$\Delta \vec{E} = \frac{\nabla \cdot \rho}{\varepsilon_0} + \nabla \times \frac{\partial \vec{B}}{\partial t} + \frac{1}{\lambda_L^2} \vec{E} \tag{9}$$

This confirms that electric fields in a superconductor should be indeed shielded over the penetration depth $\lambda_L$ similar to magnetic fields. However, following the two-fluid model, that should be only the case at T=0 K. Otherwise, the normal fluid electrons will still shield electric fields in the interior over the much shorter Thomas-Fermi length $\lambda_{TF}$. So in reality, a combination of both $\lambda_L$ and $\lambda_{TF}$ will appear that should depend on a function of the density ratio $n_s/n_n$ that does not appear to be trivial. However, as the superfluid density $n_s$ increases, the electric field penetration depth should increase as indeed experimentally observed. Unfortunately, the electric field penetration measurements on the YBCO sample[7] was done only down to 60 K. It would be worthwhile to conduct such measurements approaching 0 Kelvin to see if the electric field penetration depth also approaches the London penetration depth as it is expected from the Proca model.

Solving Equ. (9) assuming rotational symmetry with the boundary condition that the electric field at r=0 must be the external electric field $E_0$, we arrive at

$$\vec{E} = \left\{ E_0(r) \cdot e^{-\frac{r}{\lambda_L}} - \left[ \frac{\nabla \cdot \rho(r)}{\varepsilon_0} \right] \cdot \left( 1 - e^{-\frac{r}{\lambda_L}} \right) \cdot \lambda_L^2 \right\} \cdot \hat{r} + \nabla \times \frac{\partial B_z(r)}{\partial t} \cdot \hat{\theta} \quad . \tag{10}$$

Whereas the external electric field $E_0$ is exponentially shielded, there exists the possibility of internal electric fields based on a gradient of charge and on a time-varying magnetic field – but only outside the London penetration depth. Only by omitting the radiation term and again without the knowledge of a massive photon, we can again extract a London-type equation for the electric field from Equ. (9) that has to be implemented in standard Maxwell equations as

$$\frac{\nabla \rho(r)}{\varepsilon_0} \cdot \left(1 - e^{-\frac{r}{\lambda_L}}\right) = \frac{1}{\lambda_L^2} E_0(r) \quad , \tag{11}$$

which is similar to the London brother's initial guess[1] and the approach from Hirsch[5] only outside of the penetration depth $\lambda_L$.

### 3. ELECTRODYNAMICS IN NON-STATIONARY SUPERCONDUCTORS

As a well-known example, we take the case of a superconductor rotating with angular velocity $\omega$. Because superfluid electrons are friction-free, the current from the rotating lattice is equivalent to a negative Cooper-pair current. That leads to the following transformations for the velocity and the electric field

$$v'_s = v_s - v_\Omega \tag{12}$$

$$\vec{E}' = \vec{E} - \frac{1}{2}\left[(\vec{\omega} \times \vec{r}) \times \vec{B}\right] \quad . \tag{13}$$

In the Lorentz force term of Equ. (13) we had to take into account that a Cooper-pair consists of 2 charges. The resulting electric and magnetic fields in a superconductor can now readily be calculated. Using the current in Equ. (12) in the Proca equation (4), we get

$$\nabla \times \vec{B} = -\mu_0 \vec{j}_\Omega + \frac{1}{c^2}\frac{\partial E}{\partial t} - \frac{1}{\lambda_L^2}\vec{A} \tag{14}$$

Taking again the curl of Equ. (14) and omitting the radiation term, we can solve for the one-dimensional case assuming rotational symmetry with the usual boundary condition that the magnetic field at r=0 must be equal to the external field $B_0$

$$B_z(r) = B_0 \cdot e^{-\frac{r}{\lambda_L}} - \frac{2m}{e}\omega \cdot \left(1 - e^{-\frac{r}{\lambda_L}}\right) \cdot \hat{z} \quad , \tag{15}$$

where the first part is the usual Meissner term and the second part the so-called London moment. Also here we see that in Proca equations, the London moment arises naturally from the massive photon and not from electrons lagging behind the lattice within the penetration depth which is the classical explanation[16]. This interpretation was also recently challenged by Hirsch[13] who showed that this is not compatible with the Lorentz force acting of the Cooper-pairs. The Proca model completely avoids those problems as the Cooper-pairs stay at rest.

The effect of a rotating superconductor can be easily computed by adding the term in Equ. (13) to our electric field solution in Equ. (10)

$$\vec{E} = \left\{ E_0(r) \cdot e^{-\frac{r}{\lambda_L}} - \left[\frac{\nabla \cdot \rho(r)}{\varepsilon_0}\right] \cdot \left(1 - e^{-\frac{r}{\lambda_L}}\right) \cdot \lambda_L^2 \right\} \cdot \hat{r} + \nabla \times \frac{\partial B_z(r)}{\partial t} \cdot \hat{\theta} - \frac{1}{2} \nabla \cdot [(\vec{\omega} \times \vec{r}) \times \vec{B}] \quad . \tag{16}$$

Combining Equs. (14-16) and assuming no time-varying fields as well as a homogenous charge distribution, we get an electric field inside the superconductor as

$$E(r) = \frac{m}{e} \dot{\omega} r \cdot \hat{\theta} + \frac{m}{e} \omega^2 r \cdot \left(1 - e^{-\frac{r}{\lambda_L}}\right) \cdot \hat{r} \quad . \tag{17}$$

The resulting electric field consists of two parts. The second part can be identified to counterbalance the centrifugal force on the Cooper-pairs. This is similar to approaches from Rystaphanik[11], Gawlinksi[12] and Hirsch[13], and points towards the interior of the superconductor – but only far away from the penetration depth $\lambda_L$, as it was also pointed out by Capellmann[17]. Rystephanick[11] proposed to compute the electromagnetic forces from a rotating superconductor such that the fictitious forces on the superconducting electrons vanish. He took the Coriolis and the centrifugal force into account. However, the fictitious forces in a rotating reference frame consist of three parts: Coriolis, centrifugal and the Euler force. The first part in Equ. (17) is to compensate for this last Euler-type fictitious force. This appeared also in derivations from Geurst el at[18] and Fischer et al[19]. Note that this contribution does not increase over the penetration depth as no external boundary condition applies in this case.

**Table 1** shows a comparison of the mechanical fictitious forces, the equivalent fields generated inside the superconductor and the resulting forces that always counterbalance the mechanical ones – but only far away from the penetration depth $\lambda_L$. We can therefore conclude that in the Proca model, the Cooper-pairs are always at rest even if the lattice is rotating.

## CONCLUSIONS

A fully consistent model to study electrodynamics for superconductors in the stationary and non-stationary regime has been developed based on Proca equations and a massive photon. The analysis appears to be more complete compared to previous approaches containing radiation and proper shielding terms and also an Euler-type electric field term that is necessary to fully compensate fictitious forces in rotating superconductors. Only this last field is not affected by the London penetration depth, whereas externally applied magnetic as well as electric fields are shielded with the same London penetration depth given by the massive photon.

## ACKNOWLEDGEMENT

I would like to thank I. Vasiljevich for fruitful discussions.

**Table 1**  Comparison of fictitious mechanical forces and their equivalent field for rotating superconductors. The force is computed using the Lorentz force equation.

| Fictitious Force | Mechanical Force | Equivalent Field in Superconductor | Force in Superconductor |
|---|---|---|---|
| Coriolis | $-2m\omega v \cdot \hat{\theta}$ | $B(r) = -\dfrac{2m}{e}\omega \cdot \left(1 - e^{-\frac{r}{\lambda_L}}\right) \cdot \hat{z}$ | $2m\omega v \cdot \left(1 - e^{-\frac{r}{\lambda_L}}\right) \cdot \hat{\theta}$ |
| Centrifugal | $-m\omega^2 r \cdot \hat{r}$ | $E(r) = \dfrac{m}{e}\omega^2 r \cdot \left(1 - e^{-\frac{r}{\lambda_L}}\right) \cdot \hat{r}$ | $m\omega^2 r \cdot \left(1 - e^{-\frac{r}{\lambda_L}}\right) \cdot \hat{r}$ |
| Euler | $-m\dot{\omega}r \cdot \hat{\theta}$ | $E(r) = \dfrac{m}{e}\dot{\omega}r \cdot \hat{\theta}$ | $m\dot{\omega}r \cdot \hat{\theta}$ |